# A FAST, ACCURATE TWO-STEP LINEAR MIXED MODEL FOR GENETIC ANALYSIS APPLIED TO REPEAT MRI MEASUREMENTS

Qifan Yang[1,4], Gennady V. Roshchupkin[2], Wiro J. Niessen[2], Sarah E. Medland[3], Alyssa H. Zhu[1], Paul M. Thompson[1], and Neda Jahanshad[1]*

1. Imaging Genetics Center, Stevens Neuroimaging and Informatics Institute, Keck School of Medicine, University of Southern California, Marina del Rey, California, 90292, USA;

2. Department of Medical Informatics, Radiology & Nuclear Medicine, Erasmus MC, Rotterdam, The Netherlands;

3. Queensland, Institute of Medical Research, Berghofer Medical Research Institute, Royal Brisbane Hospital, Brisbane, QLD 4029 Australia;

4. Computational Biology and Bioinformatics, Department of Biological Sciences, University of Southern California, California, 90089, USA;

*corresponding author email: neda.jahanshad@usc.edu

**Abstract:** Large-scale biobanks are being collected around the world in efforts to better understand human health and risk factors for disease. They often survey hundreds of thousands of individuals, combining questionnaires with clinical, genetic, demographic, and imaging assessments; some of this data may be collected longitudinally. Genetic associations analysis of such datasets requires methods to properly handle relatedness, population structure and other types of biases introduced by confounders. Most popular and accurate approaches rely on linear mixed model (LMM) algorithms, which are iterative and computational complexity of each iteration scales by the square of the sample size, slowing the pace of discoveries (up to several days for single trait analysis), and, furthermore, limiting the use of repeat phenotypic measurements. Here, we describe our new, non-iterative, much faster and accurate Two-Step Linear Mixed Model (Two-Step LMM) approach, that has a computational complexity that scales linearly with sample size. We show that the first step retains accurate estimates of the heritability (the proportion of the trait variance explained by additive genetic factors), even when increasingly complex genetic relationships between individuals are modeled. Second step provides a faster framework to obtain the effect sizes of covariates in regression model. We applied Two-Step LMM to real data from the UK Biobank, which recently released genotyping information and processed MRI data from 9,725 individuals. We used the left and right hippocampus volume (HV) as repeated measures, and observed increased and more accurate heritability estimation, consistent with simulations.

*Keywords:* SNP, Heritability, Repeated Measures, Empirical Kinship, Imaging Genetics, Genome Wide Association Studies (GWAS), Big Data, ENIGMA

## 1. Introduction

Genome-wide association studies (GWAS) human biological traits can enable access to the underlying complex genetics architecture, revealing important insights into genetic risk factors for disease and the underlying biological processes. However, large-scale populations on the order of tens to hundreds of thousands of individuals are often needed to reliably detect the relatively small effects of single nucleotide polymorphisms. This need for large sample sizes has launched widespread interest in biobank studies that collect data from large populations that may include twins or extended families; therefore, some degree of known or cryptic relatedness may exist between individuals included in the study. In order to include all participants while accurately accounting for the fact that samples may not be truly independent, which may lead to inflation of statistical results, a linear mixed model (LMM) [1] is often used to improve association power and accuracy.

Despite the growing development for better LMM methods, they still have limitations. Factored spectrally transformed linear mixed models [2], and genome-wide efficient mixed-model association [3] are two widely used exact models for LMM. However, these iterative optimization methods are computationally intensive, on the order of $O(N^2)$ for each iteration. Bayesian mixed models [4], have been proposed to overcome some of this complexity, and each iteration takes on the order of $O(N)$; yet, these methods do not estimate narrow-sense heritability, and they require the sample size to be sufficiently large to attain model accuracy. Because of the iterative optimization strategy, the computation time for most existing methods increases rapidly as the sample sizes become larger, to ensure convergence. This all makes them unfeasible for scaling on large datasets or analyzing many phenotypes, which is essential requirement for such methods in coming biobank era.

Fast methods have been motivated by large-scale genetics problems [5, 6, 7], yet these methods assume minimal or no genetic relationships among individuals. Here, we build on and fill a gap in existing work and develop a novel approach to avoid time-consuming iterative steps. We propose an efficient, unbiased and consistent Two-Step Linear Mixed Model method (Two-Step LMM), which not only can be used to model the complex genetic relationships among large populations, but also repeated measures from the same individuals. Further, this Two-Step method can be applied to other biological repeated measurements studies, to analyze the relationship between genetics and imaging, and to identify significant genetic loci. The Two-Step LMM consists of two steps: first to estimate heritability of the trait ($h^2$) and then the effect size of independent variables, β. In the first step, we use dimensionality reduction methods for preprocessing and moment-matching regression [6, 7] to estimate the variance components due to genetic and environmental effects, and ultimately SNP-based heritability; in the second step, we use Generalized Least Squares (GLS).

While the time complexity for dimensionality reduction is the same as Singular Value Decomposition (SVD), our method offers an efficient and non-iterative framework to analyze repeat measurements. Consider if each subject had one repeat measurement, then the sample size would double, and the computational burden for many other algorithms would quadruple, per iteration. We show our Two-Step LMM, for estimating heritability and fixed effects, is capable of modeling complex genetic relationships in population studies with relatedness and repeated measurements. This framework could dramatically reduce the estimation time for $h^2$ and β from hours or days [3, 4, 8], to around 5 minutes per trait, when analyzing big genetics and neuroimaging datasets, such as UK Biobank. Our method is motivated primarily by applications for neuroimaging genetics, where GWAS can be performed on a large number of traits, representing volume, or area of various regions, even over a million voxels within a brain image. [9, 10]. Compared with iterative methods, this non-iterative and accurate method will show much greater time improvement when running on GPU, given GPU optimizes the matrix multiplication.

## 2. Online Methods
### 2.1 *Linear Mixed Models for single measurements*
The common framework for modeling a trait $Y^*$ according to a genetic component, $g^*$, an environmental component, $e^*$, and fixed effects $X^*$ is as follows:

$$Y^* = X^*\beta^* + g^* + e^* \qquad (1)$$

Here, if there are N individuals, $Y^*$ for a single trait is an $N \times 1$ vector, $X^*$ is an $N \times k$ matrix of k fixed effects, such as age, sex, and in the case of MRI-based volumetric phenotypes, intracranial volume (ICV), which is commonly added as a covariate. Linear mixed models (LMM) are often used to model complex genetic relationships among individuals, including cryptic relatedness, which may be a common occurrence in large-scale population studies such as the UK Biobank. In LMM, the genetic effect $g^*$ and the total error, $e^*$, are often modeled as in [4]:

$$g^* \sim N(0, \sigma_g^{*2} K^*) \tag{2}$$

$$e^* \sim N(0, \sigma_e^{*2} \Sigma^*) \tag{3}$$

where $\sigma_g^{*2}$ is the additive genetic variance. $K^*$ is an $N \times N$ genetic relationship matrix (GRM), and every entry $K^*[i, j]$ corresponds to the degree of genetic similarity between subject i and subject j [1]. The total error, $e^*$, is modeled as a normal distribution, where $\sigma_e^{*2}$ is the environmental variance and $\Sigma^*$ is an $N \times N$ covariance matrix for total error.

The structure of $\Sigma^*$ is often given based on experimental design or prior information. In independent studies, individual subject errors are independently and identically distributed (i. i. d.) and, therefore, $\Sigma^*$ is an $N \times N$ identity matrix $I^*$.

## 2.2 Two-Step Linear Mixed Model for repeat measurements

Now, suppose the number of subjects is N; yet for each subject we now have more than one measurement, or repeat measurements, that we will use to model the genetic effects. For p measurements, equation (1) may now be written using matrix notation. Let Y, X and e be $Np \times 1$ vectors, representing repeated traits, fixed effects and environmental component.

$$Y = [y_{11}\ y_{12}\ \dots y_{1p}\ y_{21}\ y_{22}\ \dots y_{2p}\ \dots\ y_{N1}\ y_{N2}\ \dots y_{Np}]^T \tag{4}$$

$$X = [x_{11}\ x_{12}\ \dots x_{1p}\ x_{21}\ x_{22}\ \dots x_{2p}\ \dots\ x_{N1}\ x_{N2}\ \dots x_{Np}]^T \tag{5}$$

$$e = [e_{11}\ e_{12}\ \dots e_{1p}\ e_{21}\ e_{22}\ \dots e_{2p}\ \dots\ e_{N1}\ e_{N2}\ \dots e_{Np}]^T \tag{6}$$

Specifically, under the notations of (4), (5) and (6), suppose g, K, $\Sigma$ are $Np \times 1$ repeated genetic component, $Np \times Np$ repeated GRM and $Np \times Np$ covariance matrix for total error. The Linear mixed model with p measurements is similarly expressed as equation (1), (2) and (3)

$$Y = X\beta + g + e \tag{7}$$

$$g \sim N(0, \sigma_g^2 K) \tag{8}$$

$$e \sim N(0, \sigma_e^2 \Sigma) \tag{9}$$

the GRM for repeated measures, K is assumed to be expressed as $K = K^* \otimes \Delta$, where $K^*$ is an $N \times N$ GRM of subjects (no repeated measurements), and $\Delta$ is a $p \times p$ definite matrix with $\Delta[i, j] = 1$, for all i and j [11]. $\Sigma$ is the covariance matrix of total errors.

The matrix form of repeated LMM is mathematically equivalent to

$$y_{ij} = x_{ij}\beta + g_{ij} + e_{ij};\ i = 1, 2, \dots, N;\ j = 1, 2, \dots, p \tag{10}$$

where $y_{ij}$ is the $j^{th}$ repeated measurement for $i^{th}$ subject, $x_{ij}$ is a $1 \times k$ vector of k fixed effects, as above. $e_{ij}$ is the $j^{th}$ total error for $i^{th}$ subject. For example, $e_{ij}$ could represent the total effect of environmental effects and individual measurement errors.

In order to estimate heritability $h^2 = \sigma_g^2 / (\sigma_g^2 + \sigma_e^2)$, it is assumed that $\sigma_g^2$ and $\sigma_e^2$ may be estimated as $\widehat{\sigma_g^2}$ and $\widehat{\sigma_e^2}$, respectively, and therefore, $\widehat{h^2} = \widehat{\sigma_g^2} / (\widehat{\sigma_g^2} + \widehat{\sigma_e^2})$. However, going back to equations (8) and (9), we need to model K and $\Sigma$ appropriately for the repeated measurements.

For repeated GRM, K in (8), this requires modeling the repeated identities in the matrix $K^*$, and results in a noninvertible matrix; for repeated total effect $\Sigma$ in (9), subject specific unique environmental effects, which would be shared across repeat measurements for subject i, may need to be modeled, as opposed to solely the measurement errors, corresponding to unique errors for each measurement j. In the case of repeat measurements, therefore $\Sigma$ may not be an identity matrix. If there is a correlation between the errors of any p measures i and j, then $\Sigma[i,j] = (1-\rho)I + \rho I^* \otimes \Delta$, where I is a Np × Np identity matrix, $I^*$ is a N × N identity matrix and $\rho$ is the correlation coefficient [11]. In this paper, we focus on the Linear Mixed Model where the repeated measurement correlation coefficient $\rho$ is negligible. Future work will include a non-zero correlation, but studies have shown that this assumption performs similarly to the more complex model which accounts for correlated errors [6, 7]; in the case of our current paper, much of this shared error across intrasubject measurements may also be absorbed by the repeated measurement used as a covariate, as we will describe.

With these factors in mind, our goal is to obtain accurate and robust estimators $\widehat{h^2}$ and $\hat{\beta}$, for fixed effects, which is necessary for association studies taking into account relatedness. This includes a single SNP effect, $\widehat{\beta_{SNP}}$, and when tens of millions of SNPs are tested in genome-wide screens.

The Two-Step LMM consists two steps to estimate $h^2$ and $\beta$. In the first step, we use a dimensional reduction method to compute SVD-transformation matrices $S_1$ and U for preprocessing, and use moment matching regression to estimate $\sigma_g^2$, $\sigma_e^2$ and $h^2$; in the second, we use the Generalized Least Squares (GLS) method to estimate $\beta$. We describe these steps in Figure 1 and following sections.

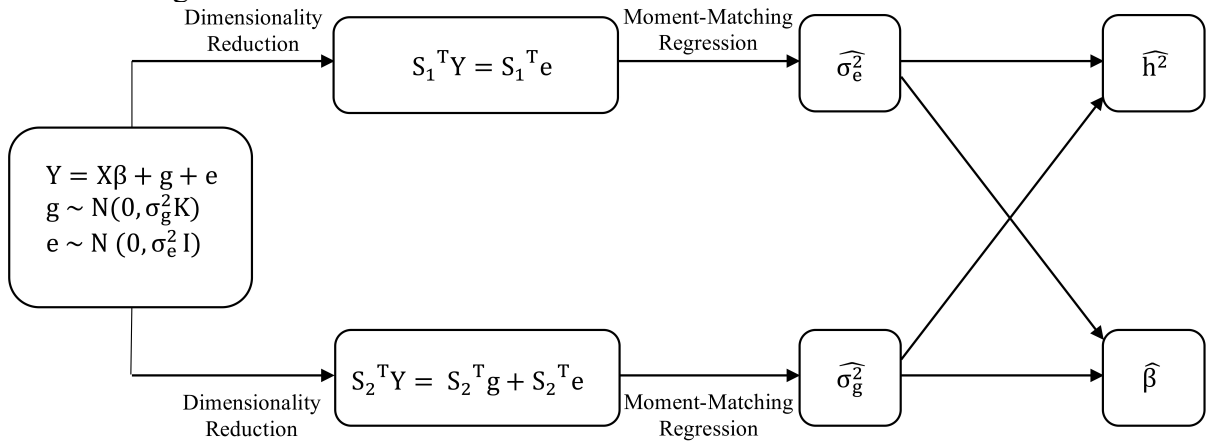

**Figure 1.** Procedure of Two-Step Linear Mix model for repeat measurements.

Here, dimensionality reduction preprocessing is necessary to obtain an efficient, unbiased and consistent $\widehat{\sigma_e^2}$. $\widehat{\sigma_g^2}$ is obtained only when $\widehat{\sigma_e^2}$ is known, and is also efficient, unbiased and consistent. Based on the statistical properties of OLS, $h^2 = \sigma_g^2 / (\sigma_g^2 + \sigma_e^2)$ is the most appropriate estimator.

### 2.2.1 *Step 1 part 1: SVD dimensionality reduction preprocessing*
First, find a SVD-transformation matrix $S_1$, such that

$$S_1^T X = 0, \ S_1^T g = 0, \ S_1^T S_1 = I \quad (11, 12, 13)$$

If the conditions stated in equations (11), (12) and (13) are satisfied, then the original linear mixed model (7) is equivalent to the following new linear mixed model, where $\sigma_e^2$ is the only unknown parameter.

$$S_1^T Y = S_1^T e \qquad (14)$$
$$S_1^T e \sim N(0, \sigma_e^2 I) \qquad (15)$$

$S_1$ can be computed using Fast Singular Value Decomposition (Fast SVD) [12].

### 2.2.2 Step 1 part 2: Estimate $\sigma_e^2$

After SVD dimensionality reduction preprocessing, we estimate $\sigma_e^2$ using moment matching regression [6, 7], which is mathematically equivalent to LD regression [13] under some constrains. SVD-transformed phenotypes are $S_1^T Y$. To estimate $\sigma_e^2$, we minimize $L_2$ loss of the 2nd moment of transformed phenotypes, and the random effect covariance of $S_1^T e$, a matrix I:

$$\widehat{\sigma_e^2} = \text{argmin} \, ||S_1^T Y Y^T S_1 - \sigma_e^2 I||_2 \qquad (16)$$

The estimation variance is approximated based on the consistency of $\widehat{\sigma_e^2}$.

### 2.2.3 Step 1 part 3: Estimate $\sigma_g^2$ and $h^2$

$\sigma_g^2$ and $h^2$ can also be estimated using SVD-moment matching regression. In order to estimate $\sigma_g^2$, we compute a SVD-transformation matrix $S_2$, such that

$$S_2^T X = 0, \, S_2^T S_2 = I \qquad (17, 18)$$

Apply the matrix U to both sides of the original linear mixed model

$$S_2^T Y = S_2^T g + S_2^T e \qquad (19)$$

where $S_2^T g \sim N(0, \sigma_g^2 S_2^T K S_2)$, and $S_2^T e \sim N(0, \sigma_e^2 I)$. Given $\widehat{\sigma_e^2}$,

$$\widehat{\sigma_g^2} = \text{argmin} \, ||S_2^T Y Y^T S_2 - \sigma_g^2 S_2^T K S_2 - \widehat{\sigma_e^2} I||_2 \qquad (20)$$

Finally, the heritability estimator is obtained through $\widehat{h^2} = \widehat{\sigma_g^2} / (\widehat{\sigma_g^2} + \widehat{\sigma_e^2})$, and variance of estimated heritability is calculated by Delta method.

### 2.2.2 Step 2: Estimate $\beta$

$$\hat{\beta} = (X^T \widehat{M}^{-1} X)^{-1} X^T \widehat{M}^{-1} Y \qquad (21)$$

in which $\widehat{M}$ is the OLS estimator of variance-covariance matrix. In particular,

$$\widehat{M} = \widehat{h^2} K + (1 - \widehat{h^2}) I \qquad (22)$$

$\hat{\beta}$ is obtained using GLS method, so $\hat{\beta}$ is also unbiased, consistent and efficient, which is an important consideration in large scale big data applications such as imaging genetics.

### 2.3 Simulation methods

To determine the accuracy and efficiency of our Two-Step LMM, we simulated several conditions similar as real UK Biobank neuroimaging and genetics data (see next section), and compared our estimates to those obtained from moment-matching regression for SNP-based Heritability Estimation (MMHE) [6, 7], whose application scope now includes repeated measurements, Repeated version of MMHE (Repeated MMHE) [7].

Suppose the total number of simulations is $N_{Sim}$, and the number of subjects is N. We perform simulation analysis for the proposed Linear Mixed models with repeat measurements:

$$Y = X\beta + g + e \qquad (7)$$

$$g \sim N(0, \sigma_g^2 K) \qquad (8)$$

$$e \sim N(0, \sigma_e^2 \Sigma) \qquad (9)$$

where here, $\Sigma$ is an identity matrix I. The number of repeated measurements p is 2. To make the simulated GRM, the matrix $K^*$, close to a real GRM in family data so as to show the estimation accuracy of our method, we first randomly generated a 2-trial binomial distributed $N \times m$ genotype matrix Z, where the number of SNPs, m is 100,000, and probability of success for each trial is 0.2. Here, the probability of success for each binomial trial corresponds to the Minor Allele Frequency (MAF). In particular, every column of Z needs to be the normalized. Then K is obtained by $K = K^* \otimes \Delta$, $K^* = Z^T Z/m$.

We modeled our simulation true values based on our real data example (see next section). Suppose fixed effects X in (7) include an intercept, age, sex, and a scaled version of intracranial volume (scaled ICV), $X = [1 \; X_{age} \; X_{sex} \; X_{scaledICV}]$.

Motivated by the distribution of data in the UK Biobank, we sampled $X_{age}$ from a normal distribution with mean 60 and variance 10. $X_{sex}$ is a binary variable, coded here as 0 for female and 1 for male. $X_{sex}$ was simulated using a discrete uniform distribution, with only two values, 0 and 1. $X_{scaledICV}$ was drawn from a normal distribution with mean 1.2 and variance 0.05, $X_{scaledICV} \sim N(1.2, 0.05)$. For robustness, we also performed a follow-up test, where $X_{scaledICV}$ was drawn from a contaminated normal distribution, in which 91% of simulated $X_{scaledICV}$ were from $N(1.2, 0.05)$ as before, while the remaining 9% are drawn from another normal distribution with a much higher variance, $N(1.2, 3)$. Therefore, $X_{scaledICV} \sim 0.91 N(1.2, 0.05) + 0.09 N(1.2, 3)$. The design of contaminated normal distribution follows from [16].

Using random sampling with replacement, 100 simulations were conducted for 10,000 subjects. For each simulated sample, both our Two-Step LMM method and MMHE [7] were run. True values of estimators are assigned as $h^2 = 0.4$, sum of errors $\sigma_g^2 + \sigma_e^2 = 170,000$, $\beta_0 = 6,600$, $\beta_{age} = -15$, $\beta_{sex} = 5$, $\beta_{scaledICV} = -1,400$.

## 2.4 Analysis with real data

The UK Biobank is a large-scale epidemiological study, collecting genotypes, brain MRIs, as well as numerous other variables from adults aged 45-75 in the UK. Information regarding brain imaging analysis from the UK Biobank is detailed in [14]. Summary volumetric data has been made available in their database and was accessed here as part of application #11559. As of July 2017, a sample size of 9,725 individuals had non-missing data on age at scan, sex, left and right hippocampal volume, and scaled intracranial volume (scaledICV). Raw genotypes were downloaded and used to create the GRM in the raremetalworker (RMW) software package (https://genome.sph.umich.edu/wiki/RAREMETALWORKER), which calculates the GRM according to [1]; we did not use the X chromosome in our GRM calculation. The resulting GRM matrix is full rank and invertible.

9,725 subjects in the UK Biobank were analyzed using our Two-Step Linear Mixed Model for a repeat analysis study of hippocampal volume heritability. To date, large-scale genome-wide association initiatives from large scale consortia, including the ENIGMA consortium, use a bilaterally averaged measure of hippocampal volume as the trait of interest [15, 16, 17]. This avoids possible left/right mismatches across different cohorts, and also reduces noise compared to running each left or right volumes separately. However, if the lateralized volumes are considered as repeated measures, the effective sample size may be higher, possibly leading to more stable estimates than averaging. To demonstrate these repeat measurements, 9,725 left and right measurements are considered separately as individual elements of trait vector Y. Fixed effects X in included the scaled ICV to control for head size, age, sex and an intercept. Due to the limited exclusion criteria, many health conditions are represented in UK Biobank; for the current study, we did not eliminate or control for any conditions and performed our estimates on the more diverse phenotypic pool.

Using this data, we 1) estimate the heritability of hippocampal volume using bilaterally averaged measures from all 9,725 subjects, resulting in 9,725 data points 2) estimate the heritability of hippocampal volume using each of the lateral measures from all 9,725 subjects as independent samples, resulting in 19,450 data points.

We also analyze heritability for all subcortical volume measures from 9,725 subjects as repeat measurements, including Caudate, Putamen, Pallidum, Thalamus, Hippocampus, Accumbens and Amygdala as shown in **Figure 2**.

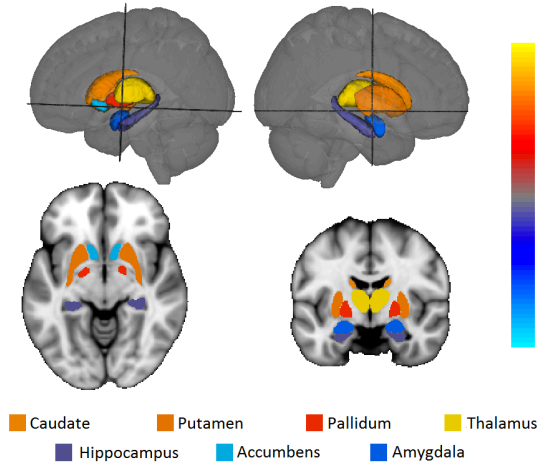

**Figure 2.** All 7 subcortical measurements from UK Biobank dataset.

## 3. Results
### 3.1 *Simulation results*
For a given simulated condition, we plot the resulting estimate of each of the 100 simulations. Simulation Set 1 (**Figure 3; Table 1**) is generated given $X_{\text{scaledICV}} \sim N(1.2, 0.05)$, while simulation Set 2 (**Figure 4; Table 2**) is the robustness test, where $X_{\text{scaledICV}} \sim 0.91 N(1.2, 0.05) + 0.09 N(1.2, 0.3)$. For Set 1 or Set 2, the same GRM (single measure or repeated measures) were run using our Two-Step LMM, and Repeated MMHE [7]. Mean, bias, standard deviation (SD) and mean squared error (MSE) are reported in **Table 1** and **Table 2** for simulated estimators, $\widehat{h^2}$, $\widehat{\sigma_g^2}$, $\widehat{\sigma_e^2}$, $\widehat{\beta_{\text{age}}}$ of our Two-Step LMM. Repeated MMHE does not explicitly provide estimates of β values, so we only report the outputs for $\widehat{h^2}$, $\widehat{\sigma_g^2}$, and $\widehat{\sigma_e^2}$.

*Simulation Set 1*: $N = 1000, N_{Sim} = 100, p = 2, X_{scaledICV} \sim N(1.2, 0.05)$

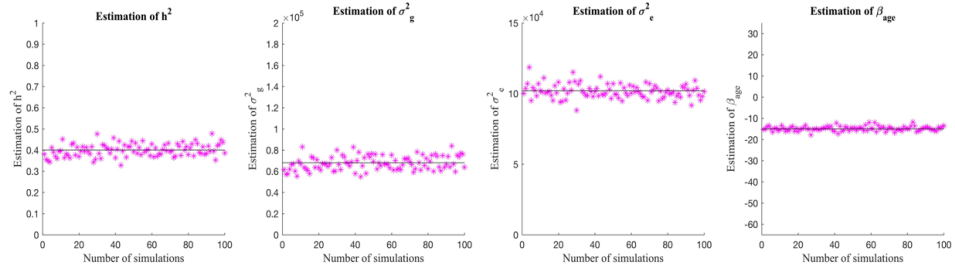

Two-Step LMM

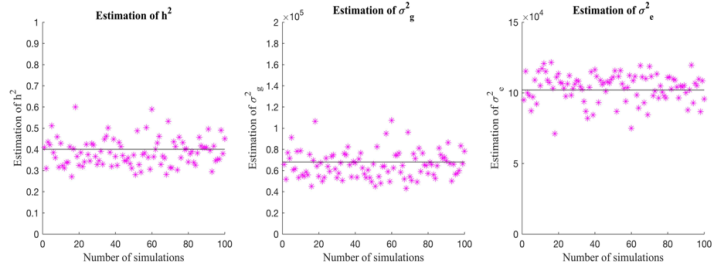

Repeated MMHE

**Figure 3.** Heritability estimate simulations ($N = 1000, N_{Sim} = 100$). Top: Two-Step Linear Mixed Model; Bottom: Repeated moment-matching heritability estimation (Repeated MMHE [7]). Columns correspond to estimations of $\widehat{h^2}, \widehat{\sigma_g^2}, \widehat{\sigma_e^2}$, and for our method, $\widehat{\beta_{age}}$.

**Table 1**. Simulation Statistics ($N = 1000, N_{Sim} = 100$) for Two-Step Linear Mixed Model and Repeated MMHE

| Estimators | True values | Mean | | Bias | | Standard Dev | | MSE | |
|---|---|---|---|---|---|---|---|---|---|
| | | Two-Step LMM | Repeated MMHE | Two-Step LMM | Repeated MMHE | Two-Step LMM | Repeated MMHE | Two-Step LMM | Repeated MMHE |
| $\widehat{h^2}$ | 0.40 | 0.40 | 0.39 | $-6.67 \times 10^{-4}$ | -0.01 | 0.03 | 0.06 | $8.84 \times 10^{-4}$ | $4.39 \times 10^{-3}$ |
| $\widehat{\sigma_g^2}$ | 68000 | 67674.47 | 65436.22 | -325.53 | -2563.78 | 6585.97 | 12768.58 | $4.35 \times 10^7$ | $1.70 \times 10^8$ |
| $\widehat{\sigma_e^2}$ | 102000 | 101591.40 | 103761.81 | -408.60 | 1761.81 | 4888.26 | 9933.70 | $2.41 \times 10^7$ | $1.02 \times 10^8$ |
| $\widehat{\beta_{age}}$ | -15.00 | -14.87 | -- | 0.13 | -- | 1.08 | -- | 1.19 | -- |

*Simulation Set 2*: $N = 1000, N_{Sim} = 100, p = 2, X_{scaledICV} \sim 0.91N(1.2, 0.05) + 0.09N(1.2, 0.3)$

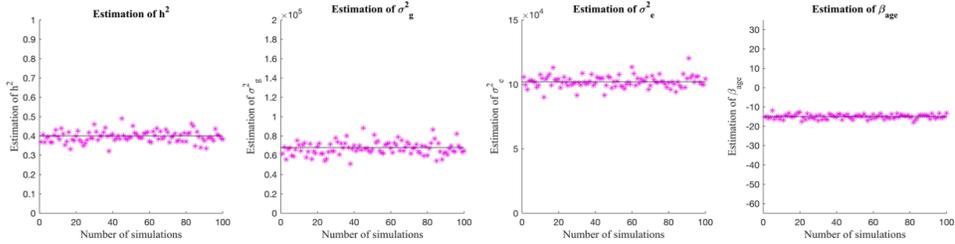

Two-Step LMM

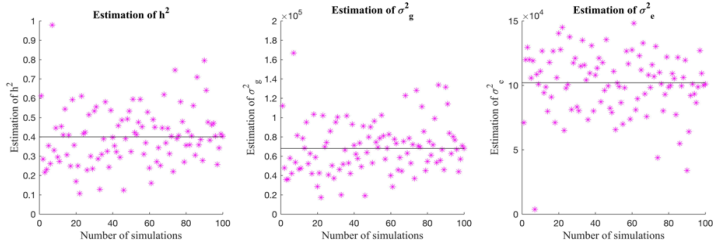

Repeated MMHE

**Figure 4.** Heritability estimate simulations (N = 1000, $N_{Sim}$ = 100, $X_{scaledICV} \sim 0.91N(1.2, 0.05) + 0.09N(1.2, 0.3)$) Top: Two-Step Linear Mixed Model; Bottom: Repeated moment-matching heritability estimation (Repeated MMHE [7]). Columns correspond to estimations of $\widehat{h^2}, \widehat{\sigma_g^2}, \widehat{\sigma_e^2}$, and for our method, $\widehat{\beta_{age}}$. The x-axis reflects the value for each of the 100 simulations;

**Table 2.** Simulation Statistics simulations (N = 1000, $N_{Sim}$ = 100, $X_{scaledICV} \sim 0.91N(1.2, 0.05) + 0.09N(1.2, 0.3)$) for Two-Step Linear Mixed Model and Repeated MMHE

| Estimators | True values | Mean | | Bias | | Standard Dev | | MSE | |
|---|---|---|---|---|---|---|---|---|---|
| | | Two-Step LMM | Repeated MMHE | Two-Step LMM | Repeated MMHE | Two-Step LMM | Repeated MMHE | Two-Step LMM | Repeated MMHE |
| $\widehat{h^2}$ | 0.40 | 0.40 | 0.40 | $-4.16 \times 10^{-3}$ | $-1.7 \times 10^{-3}$ | 0.03 | 0.15 | $9.17 \times 10^{-4}$ | 0.02 |
| $\widehat{\sigma_g^2}$ | 68000 | 67122.73 | 68794.31 | -877.27 | 794.31 | 6695.55 | 26634.63 | $4.56 \times 10^7$ | $7.10 \times 10^8$ |
| $\widehat{\sigma_e^2}$ | 102000 | 102192.35 | 101414.91 | 192.35 | -585.09 | 4456.79 | 23995.16 | $1.99 \times 10^7$ | $5.76 \times 10^8$ |
| $\widehat{\beta_{age}}$ | -15.00 | -15.00 | -- | $1.04 \times 10^{-3}$ | -- | 1.11 | -- | 1.23 | -- |

### 3.2 *Analysis with real data from the UK Biobank*

$Y^*$ is the average of left hippocampus volume and right hippocampus volume from each of the 9,725 subjects. Fixed effects $X^*$ in (1) included intercept, age, sex and scaled ICV to control for head size. Using MMHE [6] for a single (not-repeated) measurement, the heritability estimate for a single measurement is $\widehat{h^{*2}} = 0.06$ ( $\widehat{\sigma_g^{*2}} = 8027.81$ and $\widehat{\sigma_e^{*2}} = 136015.72$) which is far less than the genome-wide summary statistic heritability calculated for hippocampal volume in [18] $\widehat{h^{*2}} = 0.135$, using data from a comparable sample size (N ~ 11,600) and calculated using LD Score Regression [13] on results from the ENIGMA Consortium [16].

For hippocampus repeated measurements, we ran both our method and Repeated MMHE [7], and we compare $\widehat{h^2}, \widehat{\sigma_g^2}, \widehat{\sigma_e^2}, \widehat{\beta_0}, \widehat{\beta_{age}}, \widehat{\beta_{sex}}, \widehat{\beta_{ICV}}$ running time (implemented with MATLAB on CPU) in **Table 3**, where $\widehat{\beta_0}$ is the estimated intercept coefficient. $se(\widehat{h^2})$ is reported for Repeated MMHE [7] and Two-Step LMM, respectively. Even with repeated MMHE, which we have shown underestimates heritability, we see that this approach is far more powerful, and is capable of attributing more of the variance to additive genetic effects. Our method can capture a greater portion of the genetic variance suggested by twin and family studies to be closer to 70% [17], while also providing effect size estimates for fixed effects.

**Table 3.** Estimation results for repeated measurements of hippocampal volume in UK Biobank. The Two-Step LMM provides higher estimates of heritability as well as estimates for the beta coefficients of fixed effects.

| Estimators and Time | Repeated MMHE | Two-Step LMM |
|---|---|---|
| $\widehat{\sigma_g^2}$ | 60294.56 | 73722.11 |
| $\widehat{\sigma_e^2}$ | 130520.47 | 93669.50 |
| $\widehat{h^2}$ | 0.32 | 0.44 |
| $se(\widehat{h^2})$ | 0.07 | 0.03 |
| $\widehat{\beta_{age}}$ | -- | -14.85 |
| $se(\widehat{\beta_{age}})$ | -- | 0.45 |

Heritability results for all subcortical measurement heritability values are calculated using Two-Step LMM, as shown in **Figure 5**.

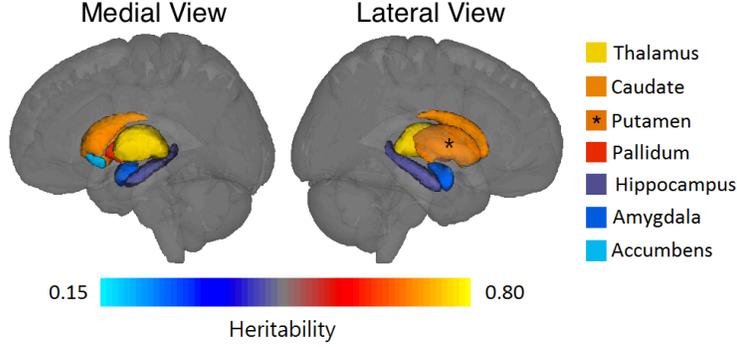

**Figure 5.** We map heritability estimates from 2StepLMM to brain mapping from UK Biobank to highlight the subcortical structure analyzed. The small asymptotic standard error of heritability (~0.02). reflects the statistical power of 2StepLMM, when applied to large-scale neuroimaging genetic datasets. Significance of age and scaled ICV are observed, while significant sex difference is not detected. CPU time per trait: 4.87 min.

## 4. Discussion

In this paper, we propose a fast, statistically efficient and accurate Two-Step Linear Mixed Model (Two-Step LMM) for population studies with cryptic relatedness and repeated measurements. Speed and accuracy are arguably the most important two criteria to evaluate a statistical model. Most iterative linear mixed models [2, 3, 4] based on Restricted Maximal Likelihood (ReML) or Bayesian methods, require many iterations to obtain accurate results.

While fast, non-iterative, linear mixed models exist, some important estimators such as $\widehat{h^2}$ or $\hat{\beta}$ for a fixed variable, may be inefficient, underestimated or inaccurate with large sampling errors, under certain conditions. This is due to the fact that these existing fast linear mixed models often ignore or just simplify the complex genetic relationship. In order to estimate β related to fixed effects such as age, sex or more importantly, SNPs, the Two-Step ProbAbel package [19] for example, first ignores genetic effects to estimate non-genetic β values. Non-repeated [6] or Repeated MMHE [7], a moment-matching heritability method works well for population data; this method gives fast and unbiased heritability estimates as shown in **Table 3**. However, for populations that include family data, the GRM deviates significantly from an identity matrix, resulting in non-normally distributed, highly correlated and heteroscedastic regression errors in moment-matching methods [20]. **Figure 3** and **Figure 4** show that if we use a model that does not support such relationships, Repeated MMHE tends to overestimate $\sigma_e^2$ and underestimate $\sigma_g^2$, resulting in underestimated $\widehat{h^2}$. In **Figure 4**, our robustness test shows that when scaled ICV contains small portion of outliers, Repeated MMHE gives $\widehat{h^2}$, $\widehat{\sigma_e^2}$, $\widehat{\sigma_g^2}$ with relatively larger standard deviations, which may explain why Repeated MMHE generates lower heritability than Two-Step LMM, since ICV data contains large portion of outliers. If we estimate β directly using results from the moment-matching regression, the bias and standard deviations of $\widehat{\sigma_e^2}$ and $\widehat{\sigma_g^2}$ will be further magnified, and the $\hat{\beta}$ will be inaccurate.

In contrast, our fast method projects the original data onto a lower dimensional space, but keeps all the necessary information for estimation. This dimensionality reduction method ensures that the estimated heritability and $\hat{\beta}$ have minimum estimation standard errors, shown in simulation studies (**Figure 3**, **Figure 4**). In **Table 3**, standard errors are reported for Repeated LMM and MMHE as in [7], indicating both methods are stable. Asymptotic standard error of $\widehat{h^2}$, is close to the simulation standard error in **Table 1** and **Table 2**. This reflects the statistical power of our Two-Step LMM, when applied on big genetics and neuroimaging data. **Table 2** and **Figure 4** show that this method is very robust, and insensitive to outliers, ensuring our heritability analysis is not biased by the outliers from the data.

Furthermore, estimates and stand error of β for age is reported in **Table 3**. As expected, main effects of age and ICV are observed, while sex difference is not detected for Hippocampus volumes in our studies. Also, lower Hippocampus heritability indicates unhealthy and healthy individual differences in our UK Biobank dataset.

In **Figure 5**, we apply Two-Step LMM to all subcortical measurements, get heritability values of Caudate, Putamen, Pallidum, Thalamus, Hippocampus, Accumbens and Amygdala. With small estimation standard errors, these heritability values are in line with the family-based heritability analysis in over 40, 000 individuals worldwide [21], indicating our fast and robust method captures the genetic architecture of the family relationship in around 10, 000 individuals.

This proposed Two-Step Linear Mixed Model has some application limitations. Most notably, the method is applied in the case where we are considering independent repeat measurements, which is regarded as the simplest case to quantify the total errors from experiments and environment. This may explain our Thalamus heritability is a little bit higher (**Figure 5**). In the more precise and complex repeated measurement model, the total error e could incorporate many additive effects, such as intrasubject and intersubject effects, household effects and measurement effects [22]. A fast, efficient and robust statistical method is still an open problem. Despite this limitation, other studies have shown that this assumption performs similarly to the more complex model which accounts for correlated errors in intrasubject measurements [6, 7]. In our case, much of this shared error across the left and right volume measurements may also be absorbed by the repeated ICV measurement fitted in X.

In future work, we will evaluate the performance of our model for genome-wide association results to identify genome-wide significant SNPs for subcortical measures or other genetic related phenotypes. Our method will also be further extended to support more complex linear mixed models, to account for additional intra-subject and inter-subject relationships, including common environment, or household effects.

**Acknowledgements**
This research has been conducted using the UK Biobank Resource *under Application Number '11559'* and was funded in part by NIH grant U54 EB020403 from the Big Data to Knowledge (BD2K) Program and R01 AG059874.